\documentclass[journal]{IEEEtran}
\usepackage{amsmath,graphicx}
\usepackage{amssymb,amsfonts,cite}
\usepackage{multirow}
\usepackage{algorithm}
\usepackage[algo2e]{algorithm2e}
\usepackage{algorithm}
\usepackage{algpseudocode}
\usepackage{xcolor}
\usepackage[mathscr]{euscript}

\begin{document}
\setlength{\belowdisplayskip}{4pt}
\title{Sensing and Classification Using Massive MIMO:\\ A Tensor Decomposition-Based Approach}

\author{B.~R.~Manoj,~\IEEEmembership{Member,~IEEE,}
        Guoda~Tian,~\IEEEmembership{Student Member,~IEEE,}
        Sara Gunnarsson,~\IEEEmembership{Student Member,~IEEE,}
        Fredrik~Tufvesson,~\IEEEmembership{Fellow,~IEEE,}  
        and~Erik~G.~Larsson,~\IEEEmembership{Fellow,~IEEE}
\vspace*{-0.5in}
\thanks{This work was supported in part by ELLIIT, Security Link and a grant from Ericsson AB. The authors also gratefully acknowledge the suggestions by Prof. Bo Bernhardsson.}
\thanks{B. R. Manoj and Erik G. Larsson are with the Department
of Electrical Engineering (ISY), Link\"{o}ping University, 58183 Link\"{o}ping, Sweden (e-mail: \{manoj.banugondi.rajashekara,  erik.g.larsson\}@liu.se).} 
\thanks{Guoda Tian, Sara Gunnarsson, and Fredrik Tufvesson are with the Department of Electrical and Information Technology, Lund University, 22100 Lund, Sweden (e-mail: \{guoda.tian, sara.gunnarsson, fredrik.tufvesson\}@eit.lth.se).}
}

\maketitle

\begin{abstract}
Wireless-based activity sensing has gained significant attention due to its wide range of applications. We investigate radio-based multi-class classification of human activities using massive multiple-input multiple-output (MIMO) channel measurements in line-of-sight and non line-of-sight scenarios. We propose a tensor decomposition-based algorithm to extract features by exploiting the complex correlation characteristics across time, frequency, and space from channel tensors formed from the measurements, followed by a neural network that learns the relationship between the input features and output target labels. Through evaluations of real measurement data, it is demonstrated that the classification accuracy using a massive MIMO array achieves significantly better results compared to the state-of-the-art even for a smaller experimental data set.
\end{abstract}

\IEEEpeerreviewmaketitle
\vspace*{-0.19in}
\section{Introduction}
\vspace*{-0.06in}
The rapid development of communication technologies over the past decade has paved the way and created a surge of interest in activity sensing applications using radio signals \cite{Qian_enabling_2018}. Exploiting characteristics of measured real-time radio channels for activity sensing is of importance for a variety of applications such as activity monitoring, security surveillance, crowd counting, elderly and children care \cite{Liu_Wi-PSG_2021, Wang_device_2017}. To realize these applications, the current state-of-the-art is 
to exploit channel state information (CSI) from existing WiFi devices that are ubiquitously available indoors \cite{Nirmal_deep_2021}. 

Early work on radio-based activity sensing was based on extraction of features from radio signals and   the use of traditional supervised machine learning (ML) algorithms to classify activities \cite{Wang_device_2017,Zhang_wivi_2020,Liu_Wi-PSG_2021}. Recently, the trend has been going towards utilizing more powerful neural network (NN)-based classifiers to further improve the classification accuracy when having multiple possible events. These classifiers in general require a huge training data set \cite{Wang_Channel_2019, Ding_wihi_2020, Zhou_FreeTrack_2020}. 
A few examples of   recent studies on    activity sensing  are   classification of different human activities \cite{Wang_device_2017}, human identity identification \cite{Ding_wihi_2020}, human tracking \cite{Zhou_FreeTrack_2020}, gesture recognition \cite{Abdelnasser_WiGest_2015},  and line-of-sight (LOS)/non line-of-sight (NLOS) identification in vehicle-to-vehicle (V2V) networks using multiple-input multiple-output (MIMO)-based systems \cite{Huang_machine_2020}. 

These applications are either using WiFi-based devices operating at a carrier frequency of 2.4~GHz (or 5~GHz), or a MIMO-based channel sounder for the V2V propagation scenario, operating at 5.9~GHz \cite{Wang_a_real_2017,Huang_machine_2020}. Current WiFi-based sensing applications use CSI measured by conventional (small-scale) MIMO systems, limiting the performance due to a smaller number of antennas, which imposes difficulties in exploiting the spatial resolution abilities. To overcome this and take activity sensing and classification one step further, we started in \cite{Manoj_Moving_2021} to consider massive MIMO for these applications. A massive MIMO array offers very high spatial resolution along with being robust to interference.
By increasing the number of antennas at the base station (BS), we can also exploit the spatial domain when extracting features from the data and hence, improve the performance. For this purpose, we conducted experiments with different activities in an indoor environment using a massive MIMO testbed operating at 3.7~GHz. While our work in \cite{Manoj_Moving_2021} focused on binary classification between static and dynamic events, we here further advance this study by instead considering the multi-class classification problem for these activities in both LOS and NLOS propagation scenarios. 

More specifically, in \cite{Manoj_Moving_2021}, we proposed a feature extraction method using principal component analysis (PCA) for the amplitude information and linear regression analysis for the phase information. 
Herein we  further improve our approach by better exploiting the phase information and the spatial domain by developing an algorithm that uses the complex correlation properties across all three domains: time, frequency, and space. Since the measured data is a tensor spanning all three domains, the dimension is large and to reduce the size, we apply a tensor decomposition-based approach to obtain   low-rank approximations. To learn the relationship between these features and their corresponding label, we propose a NN architecture with feedforward fully connected multiple layers, which is generalized for both LOS and NLOS scenarios.  
\vspace*{-0.12in}
\section{System model}
\vspace*{-0.05in}
We consider an uplink massive MIMO system with $M$ antenna elements at the BS using orthogonal frequency division multiplexing (OFDM)  with $F$ subcarriers. The multiple user equipments (UEs) utilize different but neighboring frequency subcarriers to simultaneously sound pilot signals, which are   captured by an antenna array connected to $M$ radio-frequency (RF) chains. We capture the signal over a certain observation time, recording in total $T$ snapshots of the channel. The received matrix $\mathbf{Y}_f\in \mathbb{C}^{M\times T}$, for each subcarrier index $f$ and each UE, can be written as 
\vspace*{-6pt}
\begin{equation}
    \label{receive}
    \mathbf{Y}_f = \mathbf{H}_{f} \odot\mathbf{\Gamma}_f\hspace{1pt} + \mathbf{N}_f\, ,
    \vspace*{-3pt}
\end{equation}
where snapshot $t \in [1,T]$, subcarrier $f \in [1,F]$, RF chain $m \in [1,M]$, $\odot$ denotes the Hadamard product, the complex matrix $\mathbf{H}_f\in \mathbb{C}^{M\times T}$ represents the channel, $\mathbf{\Gamma}_f\in \mathbb{C}^{M\times T}$ represents the frequency responses of the RF chains connected to the antenna array, and $\mathbf{N}_f\in\mathbb{C}^{M\times T}$ the noise matrix for subcarrier $f$, where each element of the matrix represents the additive noise to the received signal.

It is a highly challenging task to formulate a precise model of the propagation channel $\mathbf{H}_f$ since the information about geographical locations of the transmitting UEs, as well as scatters, is unknown. In addition, knowledge about the speed and directions of moving objects is not available, which imposes limitations on estimating the corresponding Doppler shifts. Furthermore, each element in the RF chain matrix $\mathbf{\Gamma}_f$, modeled as $\mathbf{\Gamma}_f(m,t) = d_m e^{j(\varphi_m - t\,\eta_{m,f})}$, introduces uncertainties in the measured channel responses, where $d_m$, $\varphi_m$, and $\eta_{m,f}$ represent the amplitude, initial phase drift, and carrier-frequency-offset (CFO), respectively, for the $f$-th subcarrier and the $m$-th RF chain at time index $t$. For the collected measurement data, we define a third-order tensor as $\mathscr{G}\in \mathbb{C}^{T \times F \times M}$, which captures the channel in the time, frequency and spatial domains. 
\vspace*{-0.12in}
\section{Algorithm for activity classification}
\vspace*{-0.04in}
In this section, we propose a novel algorithm to extract potential features from the measured channel transfer function, which is a third-order complex tensor $\mathscr{G}$. 
From $\mathscr{G}$, we extract the raw amplitude and the complex correlation across two domains by fixing the third one; thus, obtaining complex correlated tensors across all domains. Further, we utilize the real, imaginary, amplitude, and phase information of these tensors to extract features. 
The raw channel estimates from the measurement data itself could be provided to the NN model by designing a robust and complex NN to learn the statistical features of the data set. However, this approach is not feasible in our case due to the large dimensions of the measurement data. Hence, instead of utilizing the raw tensors, we further employ a tensor decomposition-based approach   in order to obtain the best low-rank approximations \cite{Kolda_tensor_2009}. 
The aim of extracting features is to later utilize these in the learning algorithm to classify activities. Our proposed feedforward NN-based architectures will be presented in Subsection \ref{FF_NN}. 
\vspace*{-0.15in}
\subsection{Tensor decomposition} \label{tensor_decomposition}
\vspace*{-0.04in}
Before describing our proposed feature extraction algorithm, we provide a brief overview of the employed tensor decomposition-based approach using canonical decomposition and parallel factors, called the CP decomposition \cite{Bergqvist_HOSVD_2010}. For any third-order real tensor ${\mathscr{H}} \in \mathbb{R}^{T \times F \times M}$, the CP decomposition can be written as a sum of $r_\mathrm{max}$ rank-one tensors \cite{Kolda_tensor_2009} as
\vspace*{-5pt}
\begin{equation}
\label{cp}
  {\mathscr{H}} \approx {\widetilde{\mathscr{H}}} = \sum_{l = 1}^{r_\mathrm{max}} \lambda_l \hspace{1pt} \mathbf{x}_l \hspace{1pt} \circ \mathbf{y}_l\hspace{1pt}
    \circ \mathbf{z}_l \, ,
 \vspace*{-4pt}
\end{equation}
where $\circ$ is the vector outer product, the factor vectors $\mathbf{x}_l \in \mathbb{R}^{T}$, $\mathbf{y}_l \in \mathbb{R}^{F}$, and $\mathbf{z}_l \in \mathbb{R}^{M}$ are normalized to unit length with weights being absorbed into $\lambda_l$, which in turn consists of the dominant eigenvalue that could be used as features. The factor matrices are expressed as ${\bf{X}} = [{\bf{x}}_1, {\bf{x}}_2, \cdots, {\bf{x}}_{r_{\mathrm{max}}}] \in \mathbb{R}^{T \times r_{\mathrm{max}}}$, ${\bf{Y}}\in \mathbb{R}^{F \times r_{\mathrm{max}}}$, and ${\bf{Z}}\in \mathbb{R}^{M \times r_{\mathrm{max}}}$, respectively. 

For calculating the CP decomposition in (\ref{cp}), we use the alternating least squares (ALS) method \cite{Kolda_tensor_2009}. 
The ALS method is used to solve linear least-squares problems by minimizing  
the mode-$n$ of  $||{\mathscr{H}}_{(n)}-{\widetilde{\mathscr{H}}}_{(n)}||$ in an iterative manner for a fixed term, where $n = \{1,2,3\}$, $||\cdot||$ is the matrix Frobenius norm, ${\mathscr{H}}_{(n)}$ and ${\widetilde{\mathscr{H}}}_{(n)}$ denote the mode-$n$ unfolding of ${\mathscr{H}}$ and ${\widetilde{\mathscr{H}}}$, respectively. 
There is no simple method to determine the rank of a tensor and for any third-order tensor ${\mathscr{H}} \in \mathbb{R}^{T \times F \times M}$, only a weak bound on the maximum rank of ${\mathscr{H}}$ is known \cite{Kolda_tensor_2009} as $\mathrm{rank}({\mathscr{H}}) \le \mathrm{min}\{TF, TM, FM\}$.  
\begin{figure}[t]
  \centering
  \centerline{\includegraphics[width=3.6in,height=0.885in]{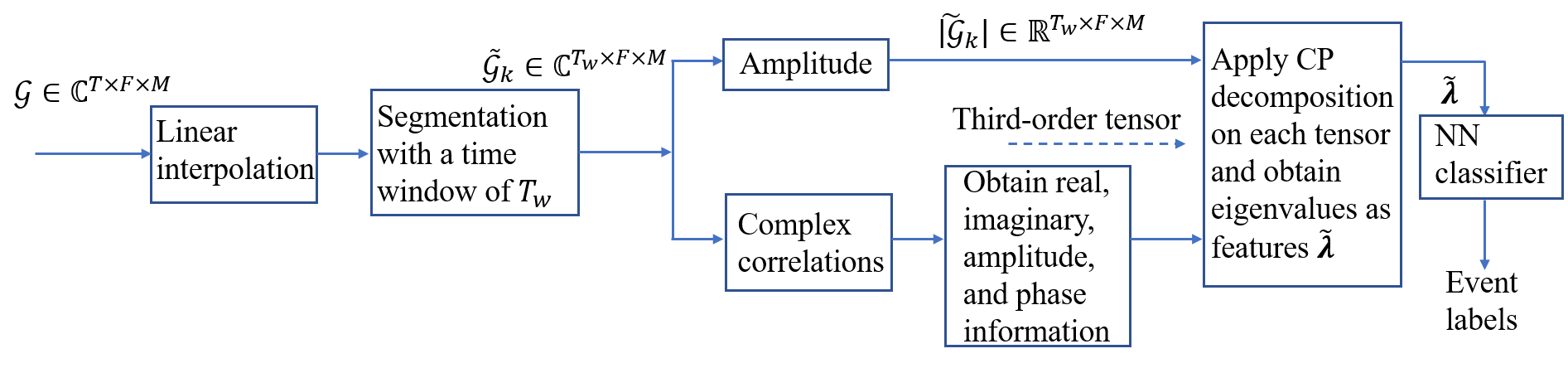}\vspace*{-0.12in}}
  \caption{Block diagram of the proposed activity sensing algorithm. 
  }
  \vspace{-20pt}
  \label{fig_block}
\end{figure}
\vspace*{-0.15in}
\subsection{Feature extraction}
\vspace{-0.05in}
The proposed flowchart for feature extraction and classification is illustrated in Fig. \ref{fig_block}. During the measurements, frame losses occur occasionally due to synchronization failures. To account for  such frame losses, we apply linear interpolation in the time domain as a first step when pre-processing the measured data,  to obtain equally spaced time domain samples.
Since the losses are intermittent, it is reasonable to expect that linear interpolation is sufficient for this purpose. 
Then, for each measurement, we perform segmentation of the received tensor $\mathscr{G}$ in the time domain, that is, a time window is used to segment the received data. The objective of the segmentation is to acquire a larger data set of shorter measurements to be used in the learning algorithm and for the performance evaluation. We choose a time window of length $T_w$ over the entire duration of the observation time $T$. The value of $T_w$ is chosen such that it is much larger than the channel coherence time so that   small changes of the time-varying correlation properties of the channels due to movement of activities are well captured. 
The time-segmented tensor data is denoted by ${\widetilde{\mathscr{G}}}_{k} \in \mathbb{C}^{T_w \times F \times M}$, where $k \in [1, T/T_w]$. 

\subsubsection{Amplitude} 
Obtaining features from the raw phase information of ${\widetilde{\mathscr{G}}}_{k}$ is challenging and may not be accurate since both the initial phase offsets ($\varphi_m$) and CFOs ($\eta_{m,f}$) of each RF chain $m$ contribute to the raw phases of the measured tensors. Furthermore, deriving an estimator to obtain those parameters is a non-trivial task. Therefore, we here only extract the feature from the raw amplitude information $|{\widetilde{\mathscr{G}}}_{k}| \in  \mathbb{R}^{T_w \times F\times M}$, where $|{\widetilde{\mathscr{G}}}_{k}|$ is the amplitude of the $k$-th time window of ${{\mathscr{G}}}$. 

\subsubsection{Complex correlations} 
To exploit the complex correlation properties of ${\widetilde{\mathscr{G}}}_{k}$, we determine the correlations as a function of two domains by fixing the third one. Thus, we evaluate the correlation functions of the channel in all three domains: time, frequency and space. For the $k$-th time window, we denote the third-order tensors $\mathscr{C}_{F,k}^M \in \mathbb{C}^{T_w\times T_w\times M}$ and  $\mathscr{C}_{T_w,k}^M\in \mathbb{C}^{F\times F\times M}$ as the complex correlation matrices for each antenna $m \in [1,M]$ across frequency and time, respectively.
More precisely, for the $m$-th antenna, the correlation matrices across frequency and time are given by  
\vspace*{-4pt}
\begin{eqnarray}\label{comp_cor_m}
\begin{aligned}
    \mathscr{C}_{F,k}^M(:,:,m) &= \widetilde{\mathscr{G}}_{k}(:,:,m)\, (\widetilde{\mathscr{G}}_{k}(:,:,m))^\mathrm{H}, \quad \mathrm{and} \\
    \mathscr{C}_{T_w,{k}}^M(:,:,m) &= (\widetilde{\mathscr{G}}_{k}(:,:,m))^\mathrm{H}\, \widetilde{\mathscr{G}}_{k}(:,:,m).
    \end{aligned}
        \vspace*{-10pt}
\end{eqnarray}
where $(\cdot)^\mathrm{H}$ is the Hermitian transpose. Similarly to \eqref{comp_cor_m}, we define $\mathscr{C}_{M,k}^F\in \mathbb{C}^{T_w \times T_w \times F}$ and $\mathscr{C}_{T_w,k}^F\in \mathbb{C}^{M\times M\times F}$ for each subcarrier $f\in[1,F]$ as tensors describing the spatial and time correlations, respectively. The tensors for each time sample $t_w \in [1,T_w]$ describing the spatial and frequency correlation matrices are denoted by $\mathscr{C}_{M,k}^{T_w}\in \mathbb{C}^{F\times F\times T_w}$ and $\mathscr{C}_{F,k}^{T_w}\in \mathbb{C}^{M\times M\times T_w}$, respectively. Following these definitions, for each subcarrier $f$, we can write 
\vspace*{-2pt}
\begin{equation}\label{comp_cor_f}
\begin{aligned}
    \mathscr{C}_{M,k}^F(:,:,f) &= \widetilde{\mathscr{G}}_k(:,f,:) \, (\widetilde{\mathscr{G}}_k(:,f,:))^\mathrm{H}, \quad \mathrm{and} \\
    \mathscr{C}_{T_w,k}^F(:,:,f) &= (\widetilde{\mathscr{G}}_k(:,f,:))^\mathrm{H} \, \widetilde{\mathscr{G}}_k(:,f,:).\\
   \end{aligned}
   \vspace*{-1pt}
\end{equation}
For each time sample $t_w$, we have 
\vspace*{-2pt}
\begin{equation}\label{comp_cor_t}
\begin{aligned}
     \mathscr{C}_{M,k}^{T_w}(:,:,t_w) &= \widetilde{\mathscr{G}}_k(t_w,:,:) \, (\widetilde{\mathscr{G}}_k(t_w,:,:))^\mathrm{H}, \quad \mathrm{and} \\
   \mathscr{C}_{F,k}^{T_w}(:,:,t_w) &= (\widetilde{\mathscr{G}}_k(t_w,:,:))^\mathrm{H} \, \widetilde{\mathscr{G}}_k(t_w,:,:).
    \end{aligned}
    \vspace*{-1pt}
\end{equation}
Each element of the tensors in (\ref{comp_cor_m})--(\ref{comp_cor_t}) is complex-valued. Thus, when utilizing the tensor decomposition-based approach discussed in Subsection~\ref{tensor_decomposition} that operates in the real domain, we first need to obtain real tensors by extracting the amplitude and unwrapped phase information from each of the tensors in (\ref{comp_cor_m})--(\ref{comp_cor_t}). Then we normalize the amplitudes and phases separately. By considering $\mathscr{C}_{F,k}^M$ in (\ref{comp_cor_m}) as an example, we denote $\mathscr{C}_{F,k}^{M,\mathrm{A}}$ and $\mathscr{C}_{F,k}^{M,\mathrm{P}}$ as the amplitude and unwrapped phase tensors, which are defined as
\vspace*{-2pt}
\begin{equation}\label{comp_corr_m_A_P}
   \begin{aligned}
     \mathscr{C}_{F,k}^{M,\mathrm{A}}(:,:,m) &= |\mathscr{C}_{F,k}^M(:,:,m)|/||\mathscr{C}_{F,k}^M(:,:,m)|| \\
   \mathscr{C}_{F,k}^{M,\mathrm{P}}(:,:,m)&= \arg(\mathscr{C}_{F,k}^M(:,:,m))/||\arg(\mathscr{C}_{F,k}^M(:,:,m))||,\hspace{-1pt}
    \end{aligned}
    \vspace*{-1pt}
\end{equation}
where $|\cdot|$ and $\arg(\cdot)$ denote magnitude and unwrapped phase, respectively. Following the same procedure as in 
(\ref{comp_corr_m_A_P}), we obtain the rest of the real tensors from (\ref{comp_cor_m})--(\ref{comp_cor_t}) with the dimension of each real tensor is same as the corresponding complex tensor. The complete list of real tensors is provided in Table~\ref{chartreal}. 

Furthermore, we also normalize the inner and outer products of (\ref{comp_cor_m})--(\ref{comp_cor_t}) with the Frobenius norm, and then extract the real, imaginary, and amplitude information from the so-obtained normalized complex tensors.  We explain this with an example: let us denote $\widetilde{\mathscr{C}}_{F,k}^M$ as the normalized complex tensor of $\mathscr{C}_{F,k}^M$, defined for each $m$ as
\vspace{-4pt}
\begin{equation} \label{comp_corr_norm_m}
    \widetilde{\mathscr{C}}_{F,k}^{M}(:,:,m) =  \mathscr{C}_{F,k}^{M}(:,:,m)/||\mathscr{C}_{F,k}^{M}(:,:,m)||, \forall \, m \,.
    \vspace{-2pt}
\end{equation}
Then the real, imaginary, and amplitude information is obtained as $\widetilde{\mathscr{C}}_{F,k}^{M,\mathrm{Re}} = \mathrm{Re}\{\widetilde{\mathscr{C}}_{F,k}^M\}$, $\widetilde{\mathscr{C}}_{F,k}^{M,\mathrm{Im}} = \mathrm{Im}\{\widetilde{\mathscr{C}}_{F,k}^M\}$, and $\widetilde{\mathscr{C}}_{F,k}^{M,\mathrm{A}} = |\widetilde{\mathscr{C}}_{F,k}^M|$, respectively. We have noticed that the phase information provided in (\ref{comp_corr_m_A_P}) is already sufficient for our classification task.  
By following the same procedure for the remaining tensors in (\ref{comp_cor_m})--(\ref{comp_cor_t}), we obtain all the real tensors as shown in Table~\ref{chartreal}, which describes the correlation properties as a function of time, frequency and space. After obtaining all the real tensors, we finally apply the CP decomposition using the ALS algorithm to achieve the best low-rank approximation for the individual real tensors and then to obtain the eigenvalues vectors ${\boldsymbol{\lambda}}_{i} \in \mathbb{R}^{r_\mathrm{max}}$ with $i\in\{1, \cdots, 31\}$, where each vector ${\boldsymbol{\lambda}}_{i}$ is sorted in the descending order. These vectors are then used as potential features in the NN learning algorithm. \vspace*{-5pt}
\begin{table}[]
\caption{List of both complex and real tensors and features}
\vspace{-5pt}
\label{chartreal}   
\renewcommand{\arraystretch}{1.4}
\centering \scalebox{0.79}{
\begin{tabular}{|l|l|l|l|l|l|} 
\hline
\begin{tabular}[c]{@{}l@{}}Complex \\ tensors\end{tabular} & \begin{tabular}[c]{@{}l@{}}Real \\ tensors\end{tabular} & Features & \begin{tabular}[c]{@{}l@{}}Complex \\ tensors\end{tabular} & \begin{tabular}[c]{@{}l@{}}Real \\ tensors\end{tabular} & Features \\ \hline
${\widetilde{\mathscr{G}}}_{k}$ & $|{\widetilde{\mathscr{G}}}_{k}|$ & ${\boldsymbol{\lambda}}_1$ & & &\\
\hline
\hline
\multirow{5}{*}{$\mathscr{C}_{F,k}^M$}                                      & $\mathscr{C}_{F,k}^{M,\mathrm{A}}$                                                      & ${\boldsymbol{\lambda}}_2$       & \multirow{5}{*}{$\mathscr{C}_{T_w,k}^M$}                                         & $\mathscr{C}_{T_w,k}^{M,\mathrm{A}}$                                                      & ${\boldsymbol{\lambda}}_7$       \\ \cline{2-3} \cline{5-6} 
                                                           & $\mathscr{C}_{F,k}^{M,\mathrm{P}}$                                                       & ${\boldsymbol{\lambda}}_3$       &                                                            & $\mathscr{C}_{T_w,k}^{M,\mathrm{P}}$                                                       & ${\boldsymbol{\lambda}}_8$       \\ \cline{2-3} \cline{5-6} 
                                                           & $\widetilde{\mathscr{C}}_{F,k}^{M,\mathrm{Re}}$                                                      & ${\boldsymbol{\lambda}}_4$       &                                                            & $\widetilde{\mathscr{C}}_{T_w,k}^{M,\mathrm{Re}}$                                                       & ${\boldsymbol{\lambda}}_9$       \\ \cline{2-3} \cline{5-6} 
                                                           & $\widetilde{\mathscr{C}}_{F,k}^{M,\mathrm{Im}}$                                                         & ${\boldsymbol{\lambda}}_5$       &                                                            & $\widetilde{\mathscr{C}}_{T_w,k}^{M,\mathrm{Im}}$                                                       & ${\boldsymbol{\lambda}}_{10}$       \\ \cline{2-3} \cline{5-6} 
                                                           & $\widetilde{\mathscr{C}}_{F,k}^{M,\mathrm{A}}$                                                        & ${\boldsymbol{\lambda}}_6$       &                                                            & $\widetilde{\mathscr{C}}_{T_w,k}^{M,\mathrm{A}}$                                                       & ${\boldsymbol{\lambda}}_{11}$       \\ \hline  \hline
\multirow{5}{*}{$\mathscr{C}_{M,k}^{F}$}                                         & $\mathscr{C}_{M,k}^{F,\mathrm{A}}$                                                      & ${\boldsymbol{\lambda}}_{12}$        & \multirow{5}{*}{$\mathscr{C}_{T_w,k}^{F}$}                                         & $\mathscr{C}_{T_w,k}^{F,\mathrm{A}}$                                                       & ${\boldsymbol{\lambda}}_{17}$       \\ \cline{2-3} \cline{5-6} 
                                                           & $\mathscr{C}_{M,k}^{F,\mathrm{P}}$                                                        & ${\boldsymbol{\lambda}}_{13}$        &                                                            & $\mathscr{C}_{T_w,k}^{F,\mathrm{P}}$                                                        & ${\boldsymbol{\lambda}}_{18}$        \\ \cline{2-3} \cline{5-6} 
                                                           & $\widetilde{\mathscr{C}}_{M,k}^{F,\mathrm{Re}}$                                                        & ${\boldsymbol{\lambda}}_{14}$        &                                                            & $\widetilde{\mathscr{C}}_{T_w,k}^{F,\mathrm{Re}}$                                                       & ${\boldsymbol{\lambda}}_{19}$        \\ \cline{2-3} \cline{5-6} 
                                                           & $\widetilde{\mathscr{C}}_{M,k}^{F,\mathrm{Im}}$                                                       & ${\boldsymbol{\lambda}}_{15}$        &                                                            & $\widetilde{\mathscr{C}}_{T_w,k}^{F,\mathrm{Im}}$                                                        & ${\boldsymbol{\lambda}}_{20}$        \\ \cline{2-3} \cline{5-6} 
                                                           & $\widetilde{\mathscr{C}}_{M,k}^{F,\mathrm{A}}$                                                       & ${\boldsymbol{\lambda}}_{16}$        &                                                            & $\widetilde{\mathscr{C}}_{T_w,k}^{F,\mathrm{A}}$                                                        & ${\boldsymbol{\lambda}}_{21}$        \\ \hline  \hline
\multirow{5}{*}{$\mathscr{C}_{M,k}^{T_w}$}                                         & $\mathscr{C}_{M,k}^{T_w,\mathrm{A}}$                                                      & ${\boldsymbol{\lambda}}_{22}$        & \multirow{5}{*}{$\mathscr{C}_{F,k}^{T_w}$}                                         & $\mathscr{C}_{F,k}^{T_w,\mathrm{A}}$                                                      & ${\boldsymbol{\lambda}}_{27}$        \\ \cline{2-3} \cline{5-6} 
                                                           & $\mathscr{C}_{M,k}^{T_w,\mathrm{P}}$                                                        & ${\boldsymbol{\lambda}}_{23}$        &                                                            & $\mathscr{C}_{F,k}^{T_w,\mathrm{P}}$                                                        & ${\boldsymbol{\lambda}}_{28}$        \\ \cline{2-3} \cline{5-6} 
                                                           & $\widetilde{\mathscr{C}}_{M,k}^{T_w,\mathrm{Re}}$                                                        & ${\boldsymbol{\lambda}}_{24}$        &                                                            & $\widetilde{\mathscr{C}}_{F,k}^{T_w,\mathrm{Re}}$                                                        & ${\boldsymbol{\lambda}}_{29}$        \\ \cline{2-3} \cline{5-6} 
                                                           & $\widetilde{\mathscr{C}}_{M,k}^{T_w,\mathrm{Im}}$                                                        & ${\boldsymbol{\lambda}}_{25}$        &                                                            & $\widetilde{\mathscr{C}}_{F,k}^{T_w,\mathrm{Im}}$                                                           & ${\boldsymbol{\lambda}}_{30}$        \\ \cline{2-3} \cline{5-6} 
                                                           & $\widetilde{\mathscr{C}}_{M,k}^{T_w,\mathrm{A}}$                                                       & ${\boldsymbol{\lambda}}_{26}$        &                                                            & $\widetilde{\mathscr{C}}_{F,k}^{T_w,\mathrm{A}}$                                                           & ${\boldsymbol{\lambda}}_{31}$        \\ \hline
\end{tabular}
}\vspace*{-0.2in}
\end{table}

\vspace*{-0.1in}
\subsection{Feedforward neural network} \label{FF_NN}
\vspace*{-0.02in}
We propose a feedforward NN architecture with fully connected layers comprising an input layer of dimension $\mathbb{R}^u$, $K$ hidden layers, and an output layer of dimension $\mathbb{R}^v$. The reason behind choosing NN instead of classical supervised ML is that, in this work, the classification problem includes different types of activities and the channel and signal models are unknown due to random Doppler shifts, hence the statistical characteristics are non-linear. We therefore  consider the NN-based approach, which has the ability to transform the input domain through a linear mapping and utilize a non-linear activation function to learn a non-linear pattern between a known input and the target output labels. 

A simple feedforward NN classifier model is considered, denoted as $f(.;{\boldsymbol{\theta}}) : {\bf{x}} \in {\cal{X}} \rightarrow {\bf{c}} \in {\cal{C}}$, where ${\boldsymbol{\theta}}$ is the model hyperparameters, ${\cal{X}} \subset \mathbb{R}^u$ is the input to the NN with $u$ as the dimension of the input, and ${\cal{C}} \subset \mathbb{R}^v$ is the label (one-hot) with $v$ being the number of classes.
The optimization loss function is denoted as ${\cal{L}}({\boldsymbol{\theta}}, {\bf{x}}, {\bf{c}} )$; for the classification task, the categorical cross-entropy between $f({\bf{x}})$ and ${\bf{c}}$ is applied to this function. More specifically, the measurements data set is defined as $\{{\mathscr{G}}_j, {y}_j\}$, $j = 1, \cdots, N$, where $N$ is the size of the data set, ${\mathscr{G}}_j \in \mathbb{C}^{T \times F \times M}$ is the $j$-th measurement data tensor of an activity, and ${y}_j$ is the corresponding label. The extracted feature set with the raw amplitude and correlation properties 
of the received tensors is given by $\widetilde{{\boldsymbol{\lambda}}} = \left[{\boldsymbol{\lambda}}_1, \cdots, {\boldsymbol{\lambda}}_{31} \right]$, where ${\boldsymbol{\lambda}}_i \in \mathbb{R}^{r_\mathrm{max}}$ with $i = \{1,\cdots,31\}$. Therefore, the input  to the NN becomes the feature set, which is defined as ${\bf{x}} = \mathrm{vec}(\widetilde{{\boldsymbol{\lambda}}})$ and the corresponding output of the NN is the one-hot encoded label ${\bf{c}}$. We propose a generalized NN architecture model for both the LOS and NLOS propagation scenarios, specifically a feedforward fully connected multi-layer NN with parameters  $64-32-32-32-5$ and corresponding activation functions after the linear mapping  $\textit{elu}-\textit{elu}-\textit{elu}-\textit{elu}-\textit{softmax}$. The hyperparameter set is designed as: optimizer = Adam, loss function = categorical cross-entropy, and learning rate = $0.001$. In the feature set $\widetilde{{\boldsymbol{\lambda}}}$, for both the LOS and NLOS scenario, we discard the largest eigenvalue from each vector  ${\boldsymbol{\lambda}}_i$, $i \in [1,31]$, since its value is high and does not distinguish across the activities, making it difficult for accurate classification. 
\vspace*{-0.12in}
\section{Measurement scenarios and setup}
\begin{figure}[t]
  \centering
  \centerline{\includegraphics[width=1.45in,height=0.75in]{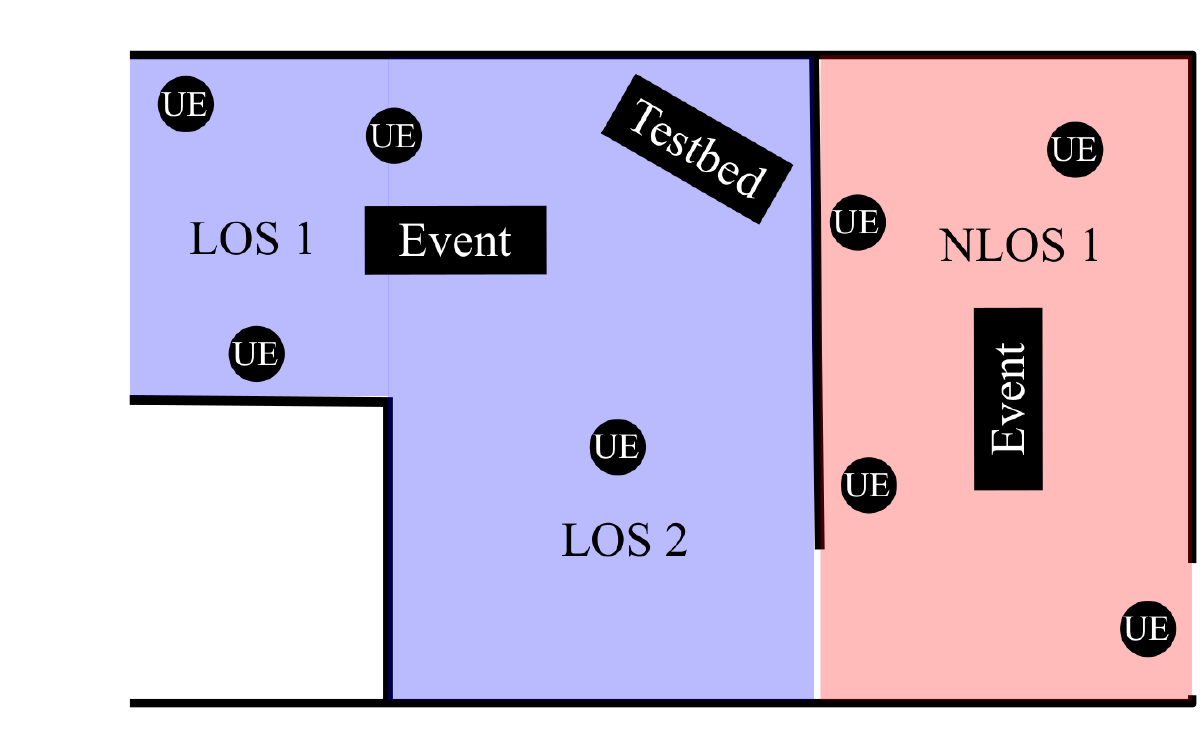}\vspace*{-0.1in}}
  \caption{Map showing the measured scenarios. The UEs were placed either in LOS~$1$ and~$2$ or in NLOS~$1$. When measuring a dynamic environment, the ``Event" rectangle was the place where the activity was performed.}
  \vspace*{-0.25in}
  \label{fig_los_nlos_map}
\end{figure}

\vspace*{-0.02in}
A measurement campaign has been carried out to collect real-time channels. Measurements were done in both LOS and NLOS as well as under static and dynamic conditions. The measurements were done in an indoor laboratory sketched in Fig.~\ref{fig_los_nlos_map}, with the Lund University massive MIMO testbed (LuMaMi) \cite{lumami_journal} acting as BS. The measurement campaign was conducted in the presence of a large number of static, scattering objects such as furniture equipment and cabinets located in the surrounding. All UEs were distributed randomly, also with placement at different heights in the LOS or the NLOS area, depending on the scenario. In case of a dynamic scenario, the activity took place in the corresponding ``Event" rectangle in the figure.

LuMaMi is based on software-defined radios (SDRs) and operates at a carrier frequency of $3.7$~GHz with $20$~MHz bandwidth in real-time. It has $100$~RF chains, each one connected to a patch antenna, building up to a rectangular array of dimensions $4\times25$ where the antennas are spaced half a wavelength apart. The antenna in the upper left corner, as seen from the front, is vertically polarized and  the polarization is alternating such that two consecutive elements always have different polarizations. The BS is collecting channel estimates for $30$~seconds per measurement, based on the received pilot signals from the UEs, which are also SDR-based. The sampling rate is $100$ Hz, resulting in  a $10$ ms time duration between adjacent samples. Each UE consists of an universal software radio peripheral (NI X310) with two RF~chains, equipped with either one or two dipole antennas. During a measurement, the channel from each UE is stored and the resulting channel transfer function will be a tensor containing $T = 3000$~snapshots, $F = 100$~frequency points and $M = 100$~BS antennas; this is referred to as one experiment.

For the activity classification, four different dynamic events were considered besides the static case. The measured events are named as follows:  (i)  $A_1-$ completely static and in the presence of a static person, static bike wheel, or static aluminium foil balloon, 
(ii) $A_2-$ waving of an aluminium foil balloon by a volunteer lying on the floor, (iii) $A_3-$ random  dancing  activities by a volunteer, (iv) $A_4-$ spinning and moving bike wheel by a volunteer lying on the floor, and (v) $A_5-$ spinning bike wheel by a volunteer without other movements. These activities were  selected to be specific movements involved in our daily life, and include the properties of stationarity ($A_1$), periodicity ($A_2$), randomness ($A_3$), as well as rotation and shifting ($A_4$ and $A_5$).  For each dynamic event and per scenario (LOS or NLOS), three measurements were done with six UEs, resulting in a total of $18$~experiments, while for the event $A_1$, we conducted $36$ experiments. The measurements of all events were conducted in turn by a few volunteers of different age groups. The experiments under LOS and NLOS scenarios were carried out at different times.
\vspace*{-0.12in}
\section{Experimental results and discussion}
\vspace*{-0.02in}
\begin{figure}[t] 
  \centering
  \centerline{\includegraphics[width=2.9in,height=1.6in]{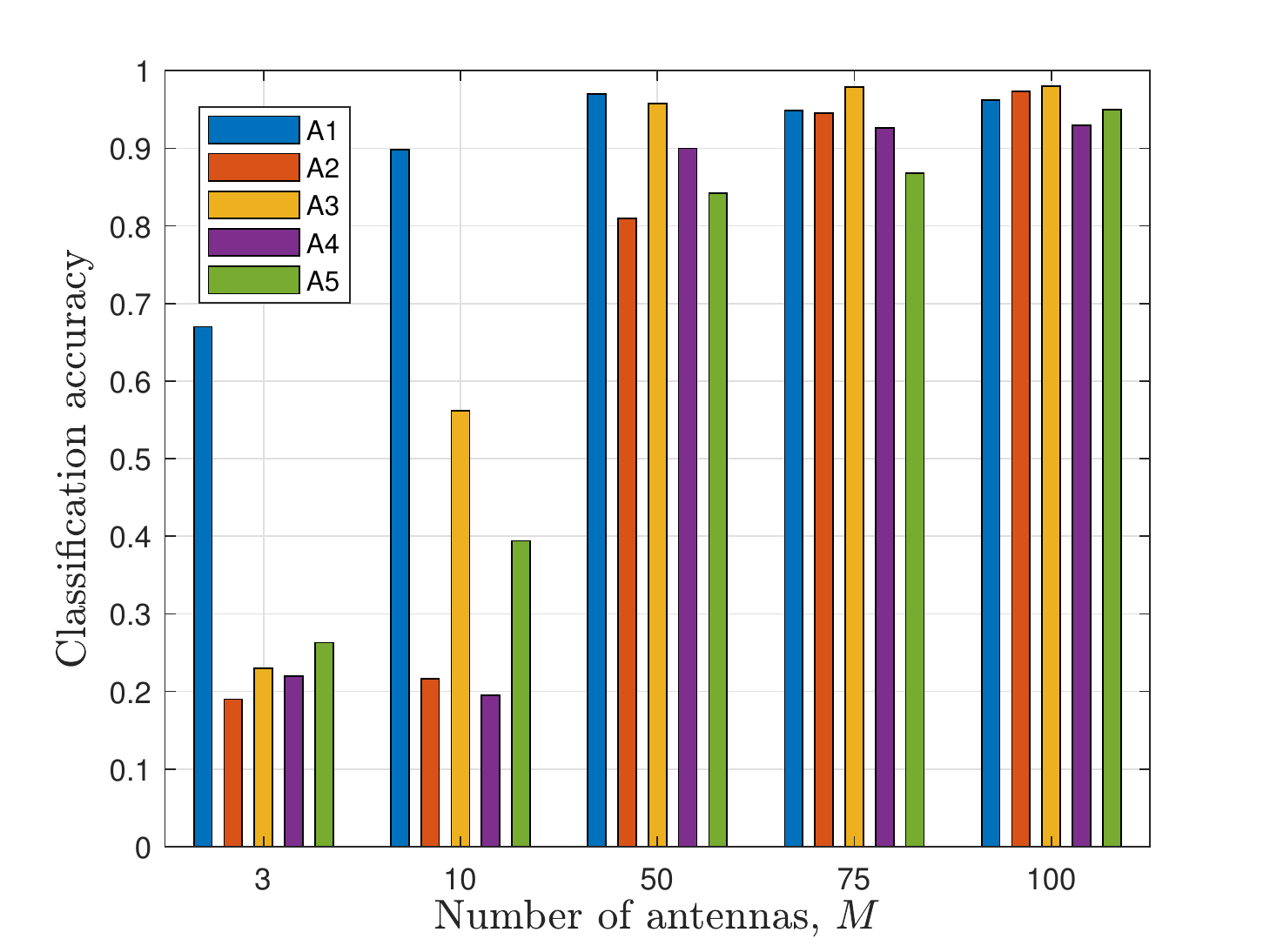}\vspace*{-0.1in}}
   \caption{LOS: Classification accuracy of activities as a function of $M$. }
  \vspace*{-0.19in}
  \label{fig:acc_los}
\end{figure}
\begin{figure}[t]
  \centering
  \centerline{\includegraphics[width=2.9in,height=1.6in]{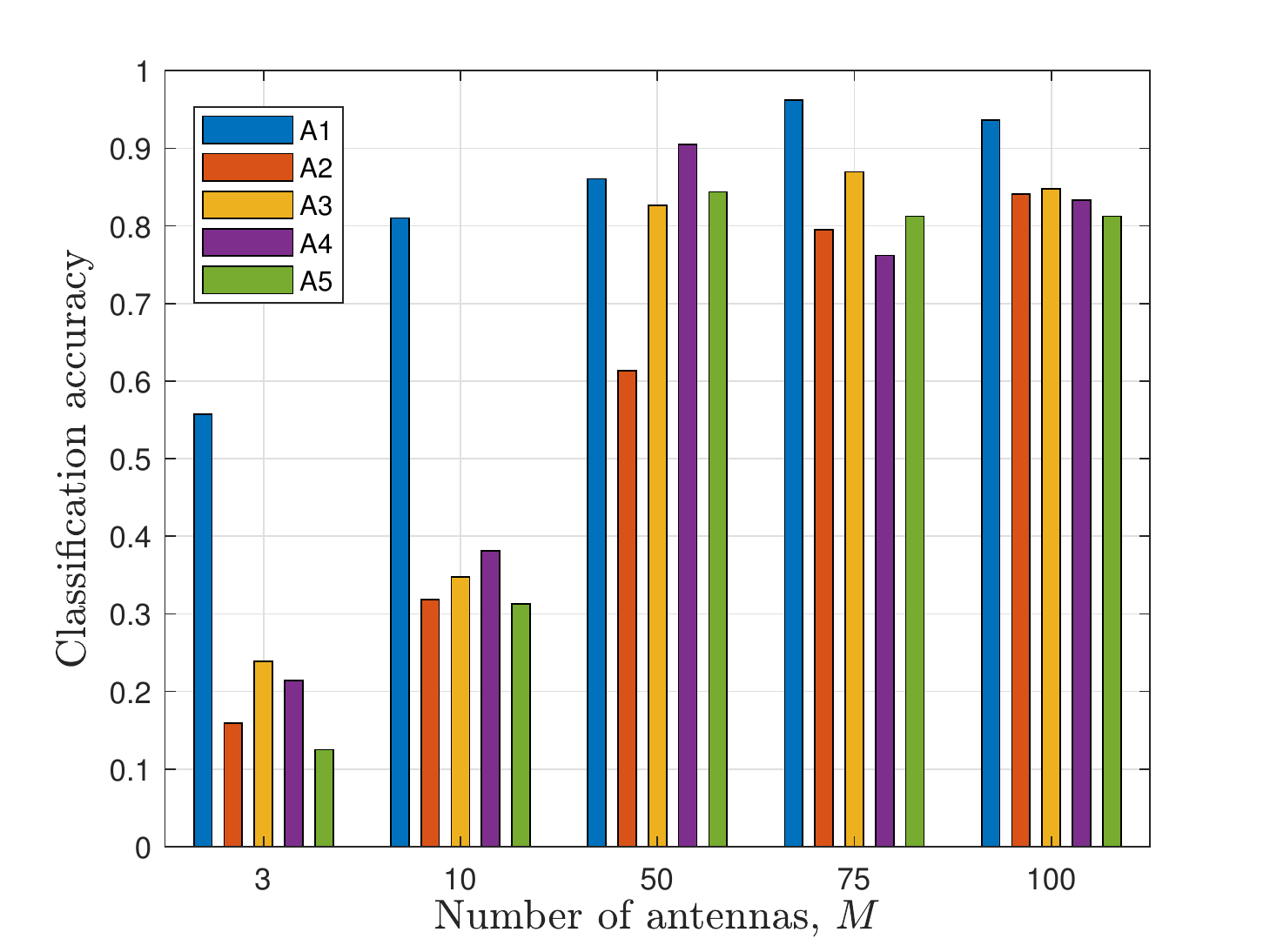}\vspace*{-0.1in}}
  \caption{NLOS: Classification accuracy of activities as a function of $M$.}
  \vspace*{-0.28in}
  \label{fig:acc_nlos}
\end{figure}

In this section, we evaluate the performance of our algorithm to classify different events under LOS and NLOS scenarios with the proposed NN model. In addition to the classification performance, the potential gain due to the extension in the spatial domain that comes with the massive MIMO array is also investigated by showing performance results for different subsets of the antenna array. In the performance evaluation, we choose ${r_\mathrm{max}} = 100$ such that the low-rank approximation gives a good fit and fulfills the aim of dimension reduction while still preserving most of the information contained in the original tensor. 
Thus, the dimension of inputs to the NN becomes $\mathbb{R}^{3069}$. 
Further, we choose $T_w = 200$ so that the data set of each experiment is increased by a factor of $T/T_w = 15$. For each dynamic event, the data set size becomes $18 \times 15 = 270$ samples, while for the static case it is $36 \times 15 = 540$ samples, where $18$ and $36$ denote the number of experiments in dynamic and static environments, respectively. Considering all the activities, in total, there are $1620$ samples. We stack the features of the events $A_1$ to $A_5$; thereby the input structure of the NN has the dimension  ${\mathbb{R}}^{1620 \times 3069}$. We then randomly split  those $1620$ samples 
into two parts, $85\%$ and $15\%$ as training and test samples, respectively. During the training phase, the NN architecture is trained with the features obtained for the events from the proposed algorithm and the corresponding target output labels.
%For each input, the proposed NN-based approach computes the probability of each event, followed by a maximum-likelihood decision. 

The classification accuracy of the activities $A_1-A_5$ with antenna array sizes ranging from $M=3$ to $M = 100$ in the LOS and NLOS scenarios is shown in Figs. 3  
and 4, respectively.   
As a special case when $M = 100$, the confusion matrices are depicted in Fig. 5. 
In the planar antenna array, the antennas are numbered row-wise starting from the upper left corner to the lower right corner. For obtaining different antenna subsets, we subsample the array in the sequence of $M = 3, 10, 50$, etc., that is, the first $3$ adjacent antennas, the first $10$ adjacent antennas, and so forth. 
\begin{figure}[t]\label{fig:confusion}
  \centerline{\includegraphics[width=3.45in,height=1.25in]{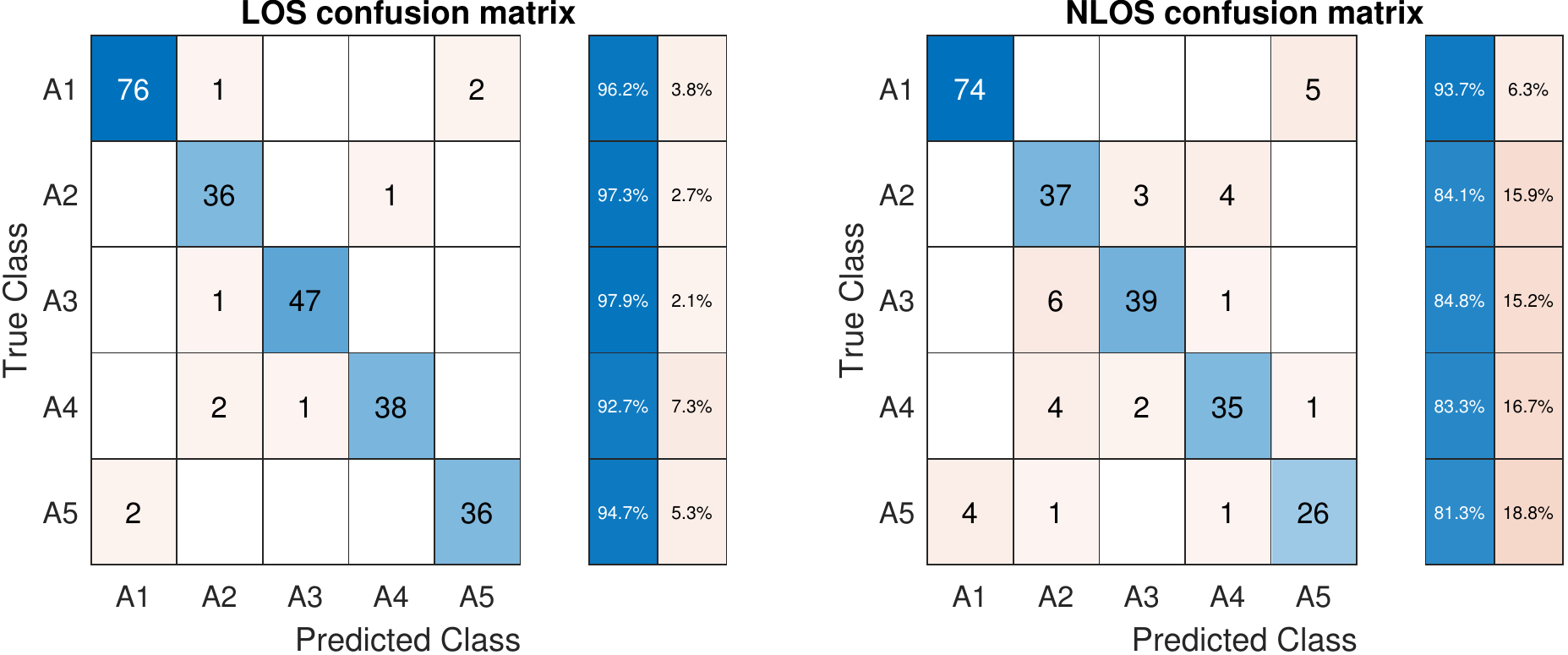} \vspace*{-0.1in}}
  \caption{Confusion matrices for LOS and NLOS scenarios when $M = 100$.}
   \vspace*{-0.25in}
\end{figure}

In both the LOS and NLOS scenarios, the classification of different activities is challenging for conventional MIMO systems (here seen for $M\leq10$), since the potential of extracting statistical characteristics embedded in the spatial domain is insufficient. However, when increasing the number of antennas, a significant performance improvement, in terms of classification accuracy, can be observed. From $M = 10$ to $M =50$, we have a substantial gain in  performance due to the aperture increase both in the horizontal and vertical directions. With $M = 100$ in the LOS scenario, the classification accuracy is $98$\%, whereas in the NLOS scenario it is $87$\%. 
For performance comparison, we implemented the PCA-based method in \cite{Manoj_Moving_2021} for the multi-class classification problem, and obtained the accuracy  83\% and 73\% for the LOS and NLOS scenarios, respectively. Thus the method proposed here outperforms that in \cite{Manoj_Moving_2021} significantly. 
In terms of complexity, we notice that it is not necessary to spend extra computational resources to achieve this accuracy gain. 
Specifically, \cite{Manoj_Moving_2021} has an overall complexity of $\mathcal{O}(TFMT_w)$ while the proposed algorithm herein has $\mathcal{O}(MFT_w\max(T_w,F,M))$, since it manages to avoid PCA computation for each time-segmented data. The classification in LOS is most likely an easier task as changes in the LOS component caused by the different activities will influence the channel more, hence making it easier to distinguish between the different activities. Also, the large-scale fading effects over the antenna array due to blockage will likely be more prominent and therefore have a larger impact on the correlation in the spatial domain; this effect is probably not that prominent in the NLOS scenario. Moreover, the UE locations could also affect the performance, since the placement of both the UE and the local scatterers around the UE influences the received signal.  However, it is challenging to predict the actual UE positions in an application scenario, since the UE distribution has a strong randomness.

To cross-validate and guarantee that our proposed algorithm learns the propagation channel characteristics rather than, e.g., gradual variations in the RF characteristics, we divide the measurement data of each experiment into two parts. Specifically, we label the first half of the entire snapshots as ``early", and the second half as ``late". For each activity, we then randomly divide $80$\% of the measurement as a training set while $20$\% as the test set, followed by training a NN with the same structure but with the size of the last layer as $2$ (i.e., binary classification).  The classification accuracy of this for each activity on the test data set shows around $50$\%, which indicates that the NN does not pick up irrelevant slow trends of the measurement data. Furthermore, the NN might (as a byproduct) learn a fingerprint associated with a specific location. However, that information is not exploited when discriminating between different activities. The NN only uses the information it learned about what the different activities look like.

\vspace*{-0.05in}
\section{Conclusions}
\vspace{-2pt}
We have investigated NN-based multi-class activity classification by utilizing baseband data from a massive MIMO array. Our proposed method is tested on data obtained from an indoor measurement campaign involving both LOS and NLOS scenarios, using the LuMaMi testbed equipped with $100$~antennas. To efficiently exploit the information embedded in the time, frequency, and spatial domains, we employed a tensor decomposition-based approach to obtain the eigenvalues as features from the three-dimensional measurement data. Furthermore, by using the proposed NN architecture, we obtained a multi-class classification accuracy that reached  98\% and 87\% in the LOS and NLOS scenarios, respectively; these numbers are significantly higher than for a system with few antennas. This showcases the potential benefits of using massive MIMO for wireless sensing applications. The results look very promising for the relatively small experimental data set we used, and further investigation is required by conducting diverse measurements in order to explore the full potential of massive MIMO for sensing applications.    
\vspace*{-0.14in}
\bibliographystyle{IEEEtran}
\bibliography{refs}
\vspace*{-0.07in}
\end{document}